 \documentclass[final,4p,times,twocolumn,authoryear,compress]{elsarticle}
\pdfoutput=1

\usepackage{amssymb}
\usepackage{amsmath}
\usepackage{amsthm}
\usepackage[]{graphicx}
\usepackage[]{subfigure}
\usepackage{tensor}
\usepackage{color}
\usepackage{cancel}
\usepackage{setspace}
\usepackage{fancyhdr}
\usepackage[bookmarks,linktocpage, colorlinks=true, plainpages = false, citecolor = blue,  linkcolor=blue, urlcolor = blue, filecolor = blue]{hyperref}

\begin{document}
\allowdisplaybreaks

\begin{frontmatter}

\title{Nucleating an Inflationary Universe: \\
Euclidean Wormholes and their No-Boundary Limit}

\author{George Lavrelashvili}
	\affiliation{Department of Theoretical Physics, A.Razmadze Mathematical Institute, at I.Javakhishvili Tbilisi State University, GE-0193 Tbilisi, Georgia}
\author{Jean-Luc Lehners}
\affiliation{Max Planck Institute for Gravitational Physics (Albert Einstein Institute), 14476 Potsdam, Germany}

\begin{abstract}
No-boundary instantons and Euclidean ``wineglass'' wormholes have both been proposed as providing suitable initial conditions for the current expanding phase of our universe, and in particular for providing conditions that are favorable to an inflationary phase. These finite action solutions have generally been regarded as unrelated, and enacting different scenarios -- in one case the creation of spacetime from nothing, and in the other up-tunneling from an Anti-de Sitter vacuum. By studying explicit solutions of both axionic and magnetic wineglass wormholes, we find that in the zero-charge limit the throat of the wormholes pinches off, leaving a no-boundary instanton that disconnects from the asymptotic Anti-de Sitter region. Thus wormholes and no-boundary instantons are part of a common family of Euclidean solutions. Along the way, we resolve the long-known puzzle that the action of wineglass wormholes can become negative. Moreover, small-charge wormholes lead to a longer inflationary phase than large-charge solutions, while no-boundary instantons dominate the probability distribution overall.
\end{abstract}




\end{frontmatter}




\section{Introduction}
\label{introduction}

Quantum gravity currently provides at least two different settings in which to study the creation of an expanding universe. The first is the no-boundary proposal, which is a prescription for describing the creation of space and time from nothing, \cite{Hartle:1983ai}. The second envisages a tunneling event from an assumed Lorentzian Anti-de Sitter ground state. The motivation for this scenario is provided by the conjectured AdS/CFT correspondence, which allows for a non-perturbative definition of quantum gravity as long as the metric is kept fixed asymptotically at its ground state value, \cite{Maldacena:1997re,Witten:1998qj}. (It remains unclear whether a third possibility, namely a bounce from a contracting phase, is allowed in quantum gravity. Additional tunneling events, of Coleman-De Luccia type, \cite{Coleman:1980aw}, may also be included in all scenarios.)

\begin{figure}
\centering
\includegraphics[scale=0.65]{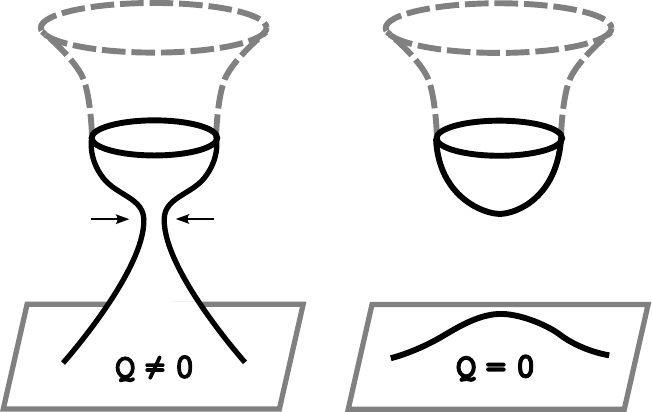}
\caption{{\it Left:} A wineglass wormhole interpolating between EAdS and a local maximum of the scale factor. Upon analytic continuation this leads to an expanding universe (indicated by the dashed lines). {\it Right:} As the axionic or magnetic charge tends to zero, the throat of the wormhole pinches off, leaving a disconnected no-boundary instanton.} \label{fig:sketch}
\end{figure}

In both settings, which so far have been considered to represent unrelated alternatives, the start of the expansion of the universe is caused by quantum events, mediated by finite action (instanton) solutions. In the case of a creation from nothing, the instantons correspond to closed, smooth geometries (with regular matter configurations) containing a single boundary, on which the field configuration is taken to correspond to the start of an inflationary expansion phase. The prototype no-boundary instanton corresponds to half of a $4$-sphere, \cite{Hartle:1983ai}. Meanwhile, up-tunneling from a Anti-de Sitter (AdS) spacetime can be achieved via Euclidean (half-)wormholes that interpolate between AdS field values and an analogous inflationary starting point. A subtlety, which was appreciated already long ago, is that in order to obtain an expanding universe after analytic continuation to Lorentzian time, one requires a wormhole of ``wineglass'' type, \cite{Lavrelashvili:1988un} (see also \cite{Rubakov:1988wx}); by this one means a geometry that first shrinks to a minimum and then expands briefly again, finally reaching a local maximum of the scale factor -- see the left panel of Fig.~\ref{fig:sketch} for a sketch. By contrast, wormholes that end at a minimum of the scale factor analytically continue to a crunching spacetime and are not suitable for our purposes.

Wineglass wormholes with asymptotically flat boundary conditions and supported by an axion field were studied numerically in \cite{Jonas:2023ipa}. Their counterparts with AdS asymptotics were conjectured to exist in \cite{Betzios:2024oli} and were proposed as initial conditions for an inflationary phase. This proposal was extended to include wormholes supported by magnetic charge and further studied in \cite{Lan:2024gnv,Betzios:2024zhf,Betzios:2026rbv}. The main reason for this proposal was the argument that, in the gravitational path integral, wineglass wormholes would have a higher weighting when interpolating to higher locations on the  potential, thus giving preference to a longer inflationary phase. This is in contradistinction to no-boundary instantons, which are known to give a higher weighting to shorter inflationary phases, \cite{Lehners:2023yrj}. It was also suggested that such wormholes could dominate over no-boundary instantons in the gravitational path integral (though it was not explained how both types of solutions could arise for a single set of boundary conditions). Crucially, no explicit wormhole solutions with AdS asymptotics were presented so far, and thus their properties were derived only by using various approximations.

We will present such solutions here and start exploring their most important properties. More in-depth studies including different scalar potentials, magnetic $SU(2)$ charge, and more exotic solutions will follow in \cite{Lavrelashvili:2026jcw}, while perturbations will be discussed in \cite{Lavrelashvili:2026nib}.

\section{Wineglass wormhole solutions}

Our first goal here is to study explicit wineglass wormhole solutions. We will simplify the analysis by restricting to metrics containing a $3$-sphere with line element $\mathrm{d}\Omega_3^2,$
\begin{align}
\mathrm{d}s^2 = \mathrm{d}\tau^2 + a^2(\tau)\mathrm{d}\Omega_3^2\,,
\end{align}
where $\tau$ denotes the Euclidean time and $a(\tau)$ the scale factor. For the sake of generality we will consider solutions with either axionic charge $Q_a$ or magnetic charge $Q_m,$ in both cases distributed spherically symmetrically. For our metric ansatz, the action then reads, \cite{Jonas:2023ipa,Marolf:2021kjc},
\begin{align}
\frac{S_E}{2\pi^2} = & \int \!\mathrm{d}\tau \left( -3 a\dot{a}^2 + \frac{1}{2}a^3 \dot\phi^2 - 3a + a^3 V + \frac{Q_a^2}{a^3} + \frac{Q_m^2}{a}\right) \nonumber \\ & - 3 a^2 \dot{a}\mid_{\tau=0}\,, \label{actionND}
\end{align}
where $\tau$ derivatives are indicated by dots. We have also included a homogeneous scalar field $\phi(\tau)$ and its potential $V(\phi).$

We are interested in half-wormhole solutions interpolating between $\tau=0$ and a large radius $\tau=\tau_{end},$ which in principle is taken to infinity at the end of the calculation. At $\tau_{end}$ we impose a Dirichlet condition by fixing the scale factor and scalar field value. In the limit $a(\tau_{end} \to \infty) \to \infty$ and $\phi(\tau_{end} \to \infty) \to \phi_{ext},$ where $\phi_{ext}$ corresponds to an extremum at negative values of the potential. At the origin, we impose a Neumann condition $\dot{a}(\tau=0)=0,\, \dot\phi(\tau=0)=0,$ which allows for a smooth analytic continuation. The surface term in \eqref{actionND} is required for a consistent variational problem, though in this case it happens to vanish since we are imposing $\dot{a}(0)=0.$

\begin{figure}
\centering
\includegraphics[scale=0.85]{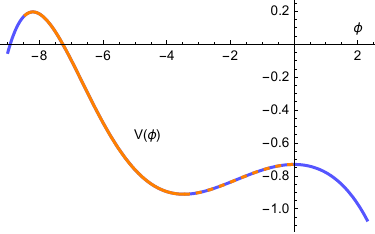}
\caption{The scalar potential $V(\phi)$ contains two negative AdS extrema and one maximum at positive values. The range of our wormhole solutions is indicated in orange -- these interpolate from either of the AdS extrema to a location just over the potential barrier provided by the positive maximum.} \label{fig:pot}
\end{figure}

The equations of motion following from the action are
\begin{align}
\ddot\phi + 3 \frac{\dot{a}}{a}\dot\phi - V_{,\phi} = 0 &\,, \\
 3\frac{\ddot{a}}{a} + \dot\phi^2 + V - 2\frac{Q_a^2}{a^6} - \frac{Q_m^2}{a^4} = 0 & \,,
\end{align}
while the associated constraint is given by
\begin{align}
3 \frac{\dot{a}^2}{a^2} - \frac{3}{a^2} =  \frac{1}{2} \dot\phi^2  - V(\phi) - \frac{Q_a^2}{a^6} - \frac{Q_m^2}{a^4}\,. \label{constraint}
\end{align}
We can use the constraint to simplify the Euclidean on-shell action which, with these boundary conditions, reduces to
\begin{align}
S^{on-shell}_E = 4\pi^2 \int \mathrm{d}\tau \left(\frac{Q_a^2}{a^3} + \frac{Q_m^2}{a} + a^3 V - 3a \right)\,.
\end{align}

With asymptotically AdS boundary conditions, the on-shell action will diverge at large volume, and this divergence must be renormalized in order to obtain physically sensible results, \cite{Balasubramanian:1998sn}. A useful strategy to achieve this is to derive the on-shell action using the Hamilton-Jacobi equation, use this in order to identify the divergent terms, and then subtract the latter in the form of appropriate counter terms, \cite{deBoer:1999tgo,Elvang:2016tzz}.

First note from \eqref{actionND} that the momenta canonically conjugate to $a, \phi$ are given by
\begin{align}
    \pi_a = -12\pi^2 a\dot{a}\,, \quad \pi_\phi = 2\pi^2 a^3 \dot\phi\,,
\end{align}
so that the Hamiltonian becomes
\begin{align}
    {\cal H} = -\frac{\pi_a^2}{24\pi^2a}+\frac{\pi_\phi^2}{4\pi^2a^3} + 2\pi^2\left( 3a-a^3V-\frac{Q_a^2}{a^3}-\frac{Q_m^2}{a}\right)\,.
\end{align}
Next we declare the on-shell action to be given by
\begin{align}
    S_{HJ}=2\pi^2 U(a,\phi)\,,
\end{align}
so that the momenta also satisfy
\begin{align}
    \pi_a = \frac{\partial S_{HJ}}{\partial a}= 2\pi^2 U_{,a}\,, \quad \pi_\phi = \frac{\partial S_{HJ}}{\partial \phi}= 2\pi^2 U_{,\phi}\,.
\end{align}
Then the Hamilton-Jacobi equation reads
\begin{align}
    -\frac{U_{,a}^2}{12a}+\frac{U_{,\phi}^2}{2a^3} +  3a-a^3V-\frac{Q_a^2}{a^3}-\frac{Q_m^2}{a}=0\,. \label{HJ}
\end{align}
We are only interested in the divergent terms in this equation as $a \to \infty.$ To this end, let us expand
\begin{align}
    U(a,\phi) = a^3 W(\phi) + a \Phi(\phi) + {\cal O}(a^{-1})\,.
\end{align}
From \eqref{HJ} we then obtain
\begin{align}
    {\cal O}(a^3): & \quad V = \frac{1}{2}W_{,\phi}^2 - \frac{3}{4} W^2\,, \label{superpot} \\
    {\cal O}(a): & \quad 3 - \frac{1}{2} W \Phi + W_{,\phi} \Phi_{,\phi} = 0\,.
\end{align}
Here we see that the relation between the functions $W$ and $V$ happens to be analogous to that between the superpotential and the potential in supergravity theories, \cite{Cremmer:1978hn}. Given this relation, it is in fact easier to start with a choice for $W(\phi),$ and to derive the scalar potential from \eqref{superpot}. Solving for $\Phi$ in closed form is however usually impossible. Asymptotically, the scalar field approaches an extremum of the potential exponentially fast. It is then sufficient to express $\Phi$ as a Taylor series, up to the highest order in $\phi$ for which $a\Phi$ still diverges (see \ref{sec:apendix1} for details). With the appropriate choice of sign for $W$ and $\Phi,$ the renormalized action is then given by $S_E^{ren} = S_E^{on-shell} + 2\pi^2 (a^3 W + a\Phi).$

\begin{figure}
\centering
\includegraphics[scale=0.45]{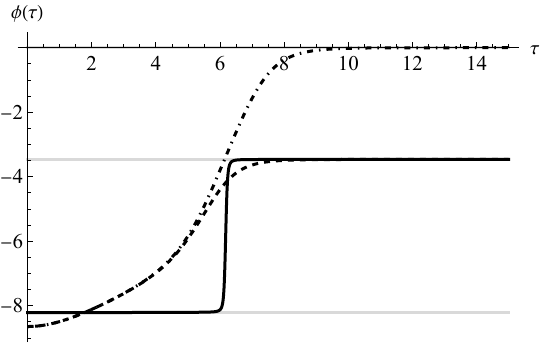}
\includegraphics[scale=0.45]{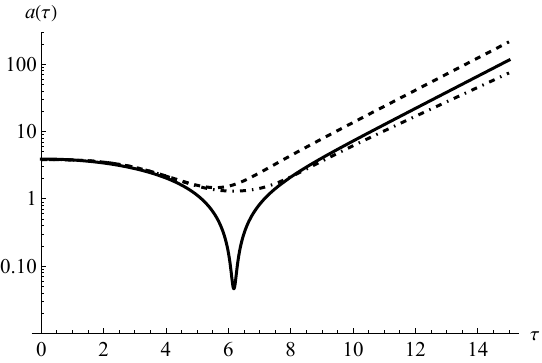}
\caption{Explicit examples of magnetically charged wineglass wormholes, with $Q_m=4$ (dot-dashed line interpolating to AdS maximum from $\phi_0=-8.64505963673769,$ dashed line interpolating to AdS minimum from $\phi_0=-8.64424753320004$) and $Q_m=0.1$ (solid line, $\phi_0=-8.21540632864282,$ interpolating to the AdS minimum). The evolution of the scalar field is shown on the left (the locations of the potential extrema are indicated by the light gray lines), and that of the scale factor on the right.} \label{fig:sol}
\end{figure}

We will present results for the case where $W$ is given by
\begin{align}
    W(\phi) = & 1+c_1 \left(\phi +\sqrt{3}-1\right)+\frac{c_2}{2} \left(\phi +\sqrt{3}-1\right)^2 \nonumber \\ & +\frac{c_3}{6} \left(\phi +\sqrt{3}-1\right)^3\,,
\end{align}
with $- c_1 = c_2 = c_3 = 1/30,$ and where we have shifted $\phi$ such that the AdS maximum lies at $\phi=0.$ The corresponding potential $V(\phi)$ is shown in Fig.~\ref{fig:pot}. The AdS minimum lies at $\phi=-2\sqrt{3}$ and the de Sitter (dS) maximum at $\phi_{max}\approx-8.21.$

We are now looking for Euclidean solutions that interpolate between either of the AdS extrema and a location $\phi_0$ at positive values of the potential. Since Euclidean evolution takes place in the inverted potential, it is clear that $\phi_0$ must be located over the top of the barrier, on the opposite side of the AdS extrema, otherwise the scalar would roll in the wrong direction. The value of $\phi_0$ cannot be determined analytically -- rather, we are using a numerical shooting technique to determine $\phi_0$ up to the required precision (typically at least a dozen significant digits). Meanwhile, the initial scale factor value $a_0$ is tied to $\phi_0$ due to the constraint \eqref{constraint}; in the case of axionic wormholes ($Q_a \neq 0, Q_m=0$) one has to solve a cubic equation, \cite{Jonas:2023ipa}, while in the case of magnetic wormholes ($Q_a=0, Q_m \neq 0$) one has to solve a quadratic equation, \cite{Lan:2024gnv}, in both cases taking the largest positive root (the smaller positive root would give a half-wormhole with a minimum of the scale factor at the origin). Requiring the roots to be real imposes an upper limit on the charge, {\it e.g.} in our case we find $Q_m \lessapprox 4.5.$

\begin{figure}
\centering
\includegraphics[scale=0.65]{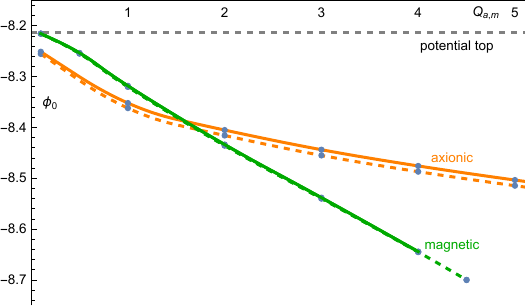}
\caption{Optimized values of the scalar field at $\tau=0,$ as a function of the axionic or magnetic charge. The solid lines are for solutions reaching the AdS minimum, and the dashed for those reaching the AdS maximum.} \label{fig:phi0}
\end{figure}

Representative explicit solutions are shown in Fig.~\ref{fig:sol}. These are supported by a magnetic charge; axionic wormholes have analogous properties (in Fig.~\ref{fig:phi0} we have plotted optimized initial data for both types, and see also \ref{sec:appendix2}). In these solutions the scalar interpolates between a location just slightly away from the potential maximum to either of the AdS extrema. When the charge is large, the interpolation is smooth (``thick wall''), while for small charges the interpolation is more abrupt (``thin wall''). As the scalar interpolates between the extrema, the scale factor, which is initially decreasing, undergoes a smooth bounce and then re-expands, asymptotically reaching the AdS solution $a(\tau) \sim \frac{1}{L}\sinh (L\tau),$ with $L=\sqrt{-3/V(\phi_{ext})}$ being the curvature length of the relevant AdS extremum. When analytically continued at the origin, these solutions lead to an expanding universe, given that $\ddot{a}<0$ becomes $a_{,tt}>0$ in Lorentzian time $t$. Thus these solutions can mediate the birth of an expanding universe by up-tunneling from an AdS vacuum. Depending on the shape of the potential outside of the region probed by the wormhole solution, an inflationary phase can follow. Moreover, as we will see shortly, wormholes that interpolate to higher locations on the potential are preferred.

\section{Small-charge limit}

\begin{figure}
\centering
\includegraphics[scale=0.7]{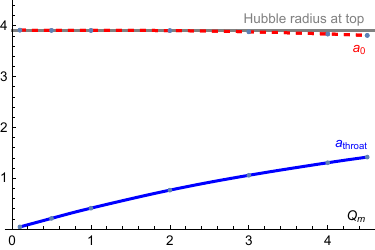}
\caption{Sizes of the ``mouth'' $a_0$ and the ``throat'' $a_{throat}$ of a family of magnetic wormhole solutions interpolating to the AdS minimum, as a function of the charge.} \label{fig:throat}
\end{figure}

It is of interest to look at the evolution of the scale factor in more detail. Fig.~\ref{fig:throat} shows the sizes of the ``mouth'' at the origin as well as the size of the ``throat'' reached at minimum scale factor. The size at the origin is fairly independent of the charge, and is given by the Hubble radius $\sqrt{-3/V(\phi_{max})}$ at the potential maximum. By contrast, the throat size depends crucially on the charge. At smaller charge values, the throat shrinks and reaches zero size in the limit of zero charge. Thus at zero charge, the throat pinches off and the topology of the solution changes. If we recall that for thin-wall solutions the scalar stays near the potential maximum more and more, then we see that at zero charge we are left with a Euclidean dS solution, with the scalar sitting at the potential maximum and the scale factor interpolating between the Hubble radius and zero size -- in other words, we are left with a disconnected no-boundary instanton! This is our key result: wineglass wormholes and no-boundary instantons are part of a common family of Euclidean solutions.

\begin{figure}
\centering
\includegraphics[scale=0.75]{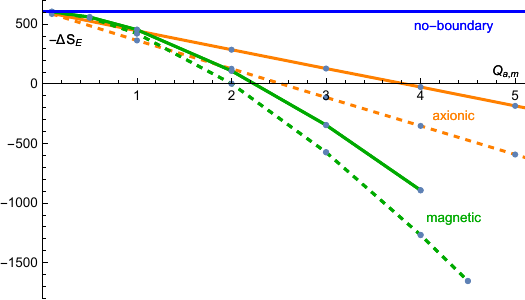}
\caption{Values of the renormalized, background-subtracted weighting $-\Delta S_E$ for the axionic and magnetic wormhole solutions of Fig.~\ref{fig:phi0}. In blue we have also plotted the weighting of the no-boundary instanton corresponding to the top of the dS maximum.} \label{fig:weight}
\end{figure}

This conclusion is reinforced by a study of the Euclidean action, see Fig.~\ref{fig:weight}. To obtain the figure, we perform two subtractions: the first is to remove the large-volume divergences in the on-shell action by adding counter terms, as described above. (We find that for solutions interpolating to the AdS maximum, we need to expand $\Phi$ up to 8th power in $\phi$.) In addition, we subtract the background value of the relevant AdS extremum, where we recall that the action of Euclidean AdS (EAdS) is given by $S_{EAdS}=-12\pi^2/V(\phi_{ext}).$ This ensures that we are considering physically meaningful quantities, see \cite{Gibbons:1976ue}. The figure shows the weighting of the solutions, which is given by $-\Delta S_E=-(S_E^{ren} - S_{EAdS}).$ In the semi-classical approximation of the gravitational path integral, the (unnormalized) probability $P$ of various solutions would then be given by the exponential of (twice) the weighting,
\begin{align}
    P \sim e^{-2\Delta S_E}\,.
\end{align}
The figure shows the same trend that was already discovered for wineglass wormholes with flat asymptotics, \cite{Jonas:2023ipa}, namely that small-charge instantons, interpolating to higher locations in the potential, come out as preferred. In fact the weighting evolves approximately linearly with the charge, especially for axionic wormholes. Also, solutions that interpolate to the AdS minimum have a higher weighting than those interpolating to a local AdS maximum. This is in contrast to the scenario proposed in \cite{Betzios:2024oli}. The most striking feature however is that the weightings of all wineglass wormhole solutions tend to the same value in the zero-charge limit, and this limit can be identified as the weighting of a no-boundary instanton sitting at the potential maximum, with
\begin{align}
    -S_E^{no-bdy} = 12\pi^2/V(\phi_{max})\,.
\end{align}
Thus, in the zero-charge limit all wineglass wormholes tend to the no-boundary instanton sitting at a local maximum of the potential -- see also the right panel in Fig.~\ref{fig:sketch}. Even background subtraction continues to work in this limit, as the full solution consists of the disconnected no-boundary and EAdS parts, $S_E^{ren}(Q=0)=S_E^{no-bdy}+S_{EAdS},$ with the action for the EAdS background being subtracted off again,
\begin{align}
    -\Delta S_E(Q=0) &= -[S_E^{ren}(Q=0)-S_{EAdS}] \nonumber \\ & = -S_E^{no-bdy}\,.
\end{align}

We have checked that entirely analogous results are obtained by using different functional forms for the potential $V$, and even hold true when the minimum of the potential is shifted to zero with the wormholes interpolating to asymptotically flat boundary conditions, see \cite{Lavrelashvili:2026jcw}.

Beyond the standard wormhole solutions, there exist oscillating solutions, \cite{Jonas:2023ipa,Lavrelashvili:2024fgu}. We do not consider them in the present work, as they are likely unstable -- analogous to the oscillatory bounces found in metastable vacuum decay, \cite{Lavrelashvili:2006cv, Battarra:2012vu}. Nevertheless, a rigorous stability analysis for the wineglass wormholes is still required; this task remains an open challenge, largely due to the persistence of the so-called 'negative mode' problem, \cite{Jonas:2023qle}.

\section{Discussion}

Our results allow us to make several observations.

First, we see that when gravity is included, the most symmetric instanton is not necessarily the most preferred one, in contrast to the case without gravity, \cite{Coleman:1977th}. Indeed, the wormholes we have considered can have a higher weighting (a lower Euclidean action) than the maximally symmetric AdS background. Our results therefore imply that the AdS vacuum is non-perturbatively unstable. This implication deserves further investigation.

Second, our results confirm that the Euclidean action of wineglass wormholes can be negative, as was already noticed long ago by \cite{Shvedov:1996hb}. However, here we see that (in a fixed potential) the action cannot be arbitrarily negative, as its value is bounded by that of the no-boundary instanton sitting at the top of the potential barrier. In this context it may be useful to recall that the sign of the action for no-boundary instantons is determined by requiring that small fluctuations about these solutions obey Gaussian rather than anti-Gaussian statistics, \cite{Lehners:2023yrj}. One may then understand the negativity of the Euclidean action for wormholes as being due to the fact that their geometry resembles the no-boundary geometry more and more when the charge is reduced.

In their original study of the decay of a false vacuum, \cite{Callan:1977pt} discuss the appearance of a volume prefactor that the path integral leads to. One might wonder what the analogue of this factor is in the present situation, especially in the zero-charge limit\footnote{We thank an anonymous referee for raising this issue.}. In \cite{Callan:1977pt}, the location of the ``bounce'' solution that mediates the tunneling can be seen as a collective coordinate, and integration over this coordinate gives rise to an infinite volume factor, which is interpreted as the volume of spacetime that is available for the new bubble to materialize. Here we may similarly ascribe a collective coordinate to the location of the stem of the wormhole, and integration over this coordinate will lead to an infinite volume factor that represents all possible wormhole locations. The quantity of physical interest is then the tunneling rate per unit spatial volume and per unit time. In the zero-charge limit, the stem shrinks and eventually pinches off. The location where the stem pinches off becomes the South Pole of a no-boundary instanton. 
Hence, in this limit, it would be desirable to interpret the spacetime volume factor as the sum over all possible locations at which the no-boundary instanton may materialize. This would ensure that the zero-charge limit remains smooth also at the level of the prefactor. 
However, once the no-boundary instanton has completely disconnected from the background geometry, whether such a summation remains justified is an open question that deserves further investigation.

And finally, comparing Figs.~\ref{fig:phi0} and \ref{fig:weight}, we conclude that wineglass wormholes obtain a higher weighting when interpolating to higher locations on the potential. However, the weighting of no-boundary solutions is even higher (our results thus only partially support the conjectures made in \cite{Betzios:2024oli}). Hence the overall probability for nucleating an expanding, inflationary universe is dominated by no-boundary instantons. Unfortunately, this implies that the old puzzle of why no-boundary instantons prefer a short inflationary phase resurfaces. Let us end with a couple of comments regarding this puzzle: If one were to stick to purely Euclidean solutions, then only the no-boundary instanton at the top of the potential would be allowed and the puzzle would disappear. Indeed, no-boundary instantons that interpolate to lower values of the potential also exist, but these are necessarily complex, \cite{Lyons:1992ua}. It seems overly restrictive however to discard all of these, \cite{Witten:2021nzp}. Rather, it will be crucial to understand which of these complex solutions should be allowed, \cite{Lehners:2021mah,Jonas:2022uqb,Lehners:2022xds,Hertog:2023vot,Lehners:2023pcn,Hertog:2024nbh}. Given the new understanding obtained in this work, it may be of interest to look for (complex?) wormholes related to the more general no-boundary instantons. This seems a promising avenue for further extending our understanding of semi-classical gravity and of the origin of our expanding universe.

\section*{Acknowledgements}
The authors thank the Albert-Einstein-Institute for kind hospitality while this work was conducted. The work of G.L. is supported in part by the Shota Rustaveli National Science Foundation of Georgia with Grant N FR-25-1194.

\appendix

\section{Counter terms} \label{sec:apendix1}

We will provide a few additional details regarding the functions $W(\phi)$ and $\Phi(\phi)$ that appear as counter terms in the combination $2\pi^2 (a^3 W + a \Phi).$ We may recall the two equations that these functions must satisfy, namely
\begin{align}
     & \quad V = \frac{1}{2}W_{,\phi}^2 - \frac{3}{4} W^2\,, \label{superpot2} \\
     & \quad 3 - \frac{1}{2} W \Phi + W_{,\phi} \Phi_{,\phi} = 0\,.\label{Phieq}
\end{align}
The first equation relates the ``superpotential'' $W$ to the scalar potential $V(\phi).$ Then assume, for the sake of argument, that $W$ is quadratic (the overall scale can be adjusted trivially),
\begin{align}
    W &= 1+w_1 \phi + \frac{w_2}{2} \phi^2\,, \\ V &= \frac{1}{2} (w_1+w_2 \phi )^2-\frac{3}{4} \left(1+ w_1 \phi +\frac{w_2 \phi ^2}{2}\right)^2\,.
\end{align}
It is straightforward to derive that there is an extremum of the potential with the value $V=-\frac{3 \left(w_1^2-2 w_2\right)^2}{16 w_2^2}$ and two degenerate extrema with $V=-w_2+\frac{1}{2}w_1^2 + \frac{1}{3}w_2^2.$ These expressions cannot accommodate the structure we are after, namely having one positive potential maximum and two extrema at negative values. This is the reason why we have to take $W$ to be (at least) cubic in $\phi$. In our case we have chosen $W$ to be of the form
\begin{align}
    W(\phi) = & 1+c_1 \left(\phi +\sqrt{3}-1\right)+\frac{c_2}{2} \left(\phi +\sqrt{3}-1\right)^2 \nonumber \\ & +\frac{c_3}{6} \left(\phi +\sqrt{3}-1\right)^3\,,
\end{align}
where we have shifted the origin of the Taylor series expansion for convenience. We would like the potential to admit a local AdS maximum at $\phi=0.$ This can be achieved by setting $W_{,\phi}(\phi=0)=0,$ which translates into the condition
\begin{align}
    c_1 = (1-\sqrt{3}) c_2 + (-2+\sqrt{3}) c_3\,.
\end{align}
We will impose this relation henceforth. It is then a rather simple matter to fix the remaining two coefficients in a suitable manner, such that we are left with one positive potential maximum and two extrema at negative values (a maximum at $\phi=0$ and a local minimum). We made the choice $- c_1 = c_2 = c_3 = 1/30,$ but other choices are equally well possible.

Our main concern lies with potential divergences as the scale factor becomes large and as the scalar field approaches one of the two extrema at negative values of the potential. Asymptotically, the scale factor behaves as
\begin{align}
    a \sim e^{\tau/L}\,, \quad \textrm{with} \quad L=\sqrt{-3/V(\phi_{ext})}\,,
\end{align}
where $\phi_{ext}$ is the value of the scalar field at a negative extremum. By expanding the scalar field equation to linear order, one can straightforwardly derive that the scalar approaches $\phi_{ext}$ as
\begin{align}
    \phi - \phi_{ext} \sim e^{-\frac{\Delta_\pm}{L} \tau} \quad \textrm{with} \quad \Delta_\pm = \frac{3}{2} \pm \sqrt{\frac{9}{4}-3\frac{V_{,\phi\phi}(\phi_{ext})}{V(\phi_{ext})}}\,.
\end{align}
There is a crucial difference between a potential maximum and a potential minimum.

Let us consider an AdS minimum first. Then $V_{,\phi\phi}>0$ at the extremum and $\Delta_+>3,\, \Delta_- <0.$ The $\Delta_-$ mode is therefore completely unstable and cannot be eliminated with counter terms. Rather, in searching for a numerical solution, the initial conditions have to be adjusted such that this mode is absent. The remaining mode, with $\Delta_+,$ is harmless and automatically renders terms of the form $a\phi^n$ convergent, for all positive $n\geq 1.$ Thus the expression for the counter term function $\Phi$ simplifies drastically, and only the constant
\begin{align}
    \Phi = \frac{6}{W(\phi_{ext})}
\end{align}
must be retained. Similarly, in the counter term $a^3W(\phi)$ one also needs to retain solely the constant term in $W.$

Now consider a local AdS maximum. In this case $V_{,\phi\phi}<0$ and we assume that $|V_{,\phi\phi}|<3|V|/4$ at the extremum (which is satisfied in our example). The dangerous mode now is the one that scales with $\Delta_-.$ Terms of the form $a\phi^n$ can diverge up to some finite power $n$ and must be removed. Hence we must solve for $\Phi$ up to that power, which for our choice of coefficients turns out to be $n=8.$ The procedure for solving \eqref{Phieq} is straightforward: one simply expands $\Phi=\sum_{k=0}^n \frac{f_k}{k!}\phi^k$ and solves \eqref{Phieq} order by order. For our choice of potential and coefficients, we find the rather unwieldy coefficients
\begin{align}
    & f_0 = \frac{540 \left(3 \sqrt{3}+94\right)}{8809},\quad f_1 =  0,\nonumber \\ &  f_2 =  -\frac{1620 \left(8971 \sqrt{3}+5076\right)}{71890249},\nonumber \\ & f_3 =  -\frac{1620 \left(35508876 \sqrt{3}+92414731\right)}{540111440737}, \nonumber \\ & f_4 =  -\frac{29160 \left(2621952327 \sqrt{3}+1030839626\right)}{3591200969460313},\nonumber \\ & f_5 =  \frac{97200 \left(60764365348870 \sqrt{3}+91389816398619\right)}{19999398198924483097}, \nonumber \\ & f_6 =  \frac{109082093236794349173000}{85457428504004316273481}\sqrt{3} \nonumber \\ & \quad + \frac{189847360281203684028000}{85457428504004316273481},\nonumber \\ & f_7 =  \frac{177300915734992337034264000}{21449814554505083384643731}\sqrt{3} \nonumber \\ & \quad + \frac{305971715864687469012221400}{21449814554505083384643731},\nonumber \\ & f_8 =  \frac{4511694631921113059364212460000}{22157658434803751136336974123}\sqrt{3} \nonumber \\ & \quad +\frac{7814979530000451109428788808000}{22157658434803751136336974123}\,. \nonumber
\end{align}
When the counter terms are added, the action converges to a finite result, which is made physically sensible by also subtracting the action of the associated EAdS background.

\section{Optimized boundary values}
\label{sec:appendix2}

For completeness, we list the optimized boundary values (at $\tau = 0$) for the wormhole solutions with unit charge:
\begin{align}
    & Q_a=1\,, \quad \textrm{to AdS minimum at $\phi=-2\sqrt{3}$}\,, \, \nonumber \\ & \phi_0 = -8.3525956\,, \quad a_0 = 3.9757434\,;  \nonumber \\
    & Q_a=1\,, \quad \textrm{to AdS maximum at $\phi=0$}\,, \, \nonumber \\ & \phi_0 = -8.3620741\,, \quad a_0 = 3.9852226\,;\qquad  \nonumber \\
    & Q_m=1\,, \quad \textrm{to AdS minimum at $\phi=-2\sqrt{3}$}\,, \, \nonumber \\ & \phi_0 = -8.3181527067\,, a_0 = 3.9066024147\,; \nonumber \\
    & Q_m=1\,, \quad \textrm{to AdS maximum at $\phi=0$}\,, \, \nonumber \\ & \phi_0 = -8.3199781398\,, a_0 = 3.9078980488\,. \nonumber
\end{align}

\footnotesize
\bibliographystyle{elsarticle-harv}
\bibliography{biblio}






\end{document}